\def\A{\leavevmode\setbox0\hbox{A}\lower1.4ex\hbox
to\wd0{\hss`}\kern-.9\wd0A}
\def\E{\leavevmode\setbox0\hbox{E}\lower1.4ex\hbox
to\wd0{\hss`\/}\kern-.9\wd0E}
\def\a{\leavevmode\setbox0\hbox{a}\lower1.4ex\hbox
to\wd0{\hss`\/}\kern-\wd0a}
\def\e{\leavevmode\setbox0\hbox{e}\lower1.4ex\hbox
to\wd0{\hss`\/}\kern-\wd0e}
\newcommand{\be}{\begin{equation}}
\newcommand{\ee}{\end{equation}}
\newcommand{\ba}{\begin{array}}
\newcommand{\ea}{\end{array}}
\newcommand{\beqn}{\begin{eqnarray}}
\newcommand{\eeqn}{\end{eqnarray}}
\newcommand{\p}{\partial}

\font\symb=msam7
\def\znakr{\raise1.5pt\hbox{\symb\char66\kern-2pt\char74}}
\def\znakl{\raise1.5pt\hbox{\symb\char73\kern-2pt\char67}}

\def\normalsize{
\setlength{\textheight}{23cm}
\setlength{\textwidth}{15cm}
\setlength{\topmargin}{-2.0cm}
\setlength{\hoffset}{-0.5cm}
\setlength{\leftmargin}{-1cm}
\setlength{\rightmargin}{2.0cm}}

\documentstyle{article}
\normalsize
\begin{document}
%%%%%%%%%%%%%%%%%%%%%%%%%%%%%%%%%%%%%%%%%%%%%%%%%%%%%%%%%%%%
%\baselineskip=24pt
%\baselineskip=20pt
%%%%%%%%%%%%%%%%%%%%%%%%%%%%%%%%%%%%%%%%%%%%%%%%%%%%%%%%%%%%
%%%%%%%%%%%%%%%%%%%%%%%%%%%%%%%%%%%%%%%%%%%%%%%%%%%%%%%%%%%%
%%%%%%%%%%%%%%%%%%%%%%%%%%%%%%%%%%%%%%%%%%%%%%%%%%%%%%%%%%%%
%%%%%%%%%% BEG\lbrack IN TITLE PAGE %%%%%%%%%%%%%%%%%%%%%%%%%%%%%%%%
%%%%%%%%%%%%%%%%%%%%%%%%%%%%%%%%%%%%%%%%%%%%%%%%%%%%%%%%%%%%
\title{On The Deser-Siegel-Townsend Notivarg}
\author{ M. Bakalarska, W. Tybor\thanks{Supported by
\L{}\'od\'z University Grant No 505/486} \\
Department of Theoretical Physics I \\
University of \L \'od\'z \\
ul.Pomorska 149/153, 90-236 \L \'od\'z , Poland}
%%%%%%%%%%%%%%%%%%%%%%%%%%%%%%%%%%%%%%%%%%%%%%%%%%%%%%%%%%%%
\date{}
\maketitle
\setcounter{section}{0}
\setcounter{page}{1}
%%%%%%%%%%%%%%%%%%%%%%%%%%%%%%%%%%%%%%%%%%%%%%%%%%%%%%%%%%%%
\begin{abstract}
The interaction of the notivarg with an external Weyl current is discussed.
The continuity equation for the Weyl current is obtained. The canonical
analysis of the theory of the notivarg interacting with the external Weyl
current is performed. The covariant propagator of the notivarg is found.
\end{abstract}
%%%%%%%%%%%%%%%%%%%%%%%%%%%%%%%%%%%%%%%%%%%%%%%%%%%%%%%%%%%%
%%%%%%%%%%% END TITLE PAGE %%%%%%%%%%%%%%%%%%%%%%%%%%%%%%%%%
%%%%%%%%%%%%%%%%%%%%%%%%%%%%%%%%%%%%%%%%%%%%%%%%%%%%%%%%%%%%
%%%%%%%%%%%%%%%%%%%%%%%%%%%%%%%%%%%%%%%%%%%%%%%%%%%%%%%%%%%%
%%%%%%%%%%%%%%%%%%%%%%%%%%%%%%%%%%%%%%%%%%%%%%%%%%%%%%%%%%%%

\newpage
\section{Introduction}

The notion of the notivarg has been introduced by Deser, Siegel and Townsend
[1] as a parallel to the Ogievetsky-Polubarinov notoph [2]. The notivarg
is a scalar particle described by the gauge theory. The notivarg field is a
twenty component tensor $K^{\mu\nu\alpha\beta}$ with symmetries of
the Riemann tensor. The Lagrangian density for the Deser-Siegel-Townsend
theory of the free notivarg is [1]
 \be
 {\cal L}_0 = - {1\over 2}(\partial_\mu K^{\mu \nu \alpha \beta} 
 \partial_\alpha{K_{\nu\kappa\beta}}^\kappa - {1\over3} \partial_\mu
 {K^{\mu\nu\alpha}}_\nu \partial_\alpha 
 {K^{\sigma\lambda}}_{\sigma\lambda}).
 \label{1.1}
 \ee
There exists another description of the notivarg [3,4] given by the
Lagrangian density
 \be
 {\cal L}_0 = -(\p_\sigma K^{\sigma\nu\alpha\beta})^2 +
 (\p_\sigma{K^{\sigma\nu\alpha}}_\nu)^2.
 \label{1.2}
 \ee
The descriptions (\ref{1.1}) and (\ref{1.2}) are not connected by the
point  transformation [3,4].
The notivarg theory based on the Lagrangian (\ref{1.2}) has been
investigated  with some details:\\
(i) the interaction of the notivarg with the external Weyl current
has been discussed in Ref. [5]; \\
(ii) the canonical analysis of the free theory and the theory of the
notivarg interacting with the external Weyl current has been performed
in Ref. [6];\\
(iii) the covariant form of the notivarg propagator has been fixed
in Ref. [7].  \\
In the present paper the similar program of investigations is performed
for the Deser-Siegel-Townsend notivarg. In Section 2 we obtain
the conservation law for the external Weyl current.
In Section 3 we carry out the canonical analysis of the theory. Its gauge
invariance is discussed in Section 4. In Section 5 we obtain the physical
Lagrangian demonstrating the pure spin-0 content of the theory.
In Section 6  we fix the covariant form of the notivarg propagator.

\section{Interaction with external Weyl current}

Let us discuss the notivarg theory in the Deser-Siegel-Townsend
description.We take into account the interaction  of the notivarg
with an external Weyl current $j^{ \mu \nu \alpha \beta }$.
The action integral has the form
 \be
 I = \int d^4 x({\cal L}_0 + {\cal L}_{int}) =
 \int d^4 x {\cal L},
 \label{2.1}
 \ee
where the free Lagrangian density ${\cal L}_0$ is given by Eq.\
(\ref{1.1})  and the interaction term is
 \be
 {\cal L}_{int} = {1\over4} j_{\mu\nu\alpha\beta}
 K^{\mu\nu\alpha\beta}= {1\over4} j_{\mu\nu\alpha\beta}
 C^{\mu\nu\alpha\beta}.
 \label{2.2}
 \ee
The 20-component field $K^{\mu\nu\alpha\beta}$ has the symmetry of the
Riemann
tensor, i.e.\\ $K^{\mu\nu\alpha\beta} = -K^{\nu\mu\alpha\beta}
= K^{\alpha\beta\mu\nu}$, $\varepsilon_{\mu\nu\alpha\beta}
K^{\mu\nu\alpha\beta}=0$;\\
the 10-component current $j^{\mu\nu\alpha\beta}$ has the symmetry of
the Weyl tensor, i.e.\ it is the Riemann tensor with
${j^{\mu\nu\alpha}}_\nu=0$ . $C^{\mu\nu\alpha\beta}$ is the Weyl
part of $K^{\mu\nu\alpha\beta}$ (see Appendix I).  \\
The free part of the action \\
 \be
 I_0 = \int d^4 x {\cal L}_0
 \ee
is invariant under the following gauge transformations
 \begin{eqnarray}
 \delta K^{\mu\nu\alpha\beta} & = &
 [{\varepsilon ^{\mu\nu}}_{\lambda\eta}
 \p^\lambda  (\p^\alpha \omega ^{\eta\beta} - \p^\beta
 \omega ^{\eta \alpha }) + {\varepsilon^{\alpha\beta}}_{\lambda\eta}
 \p^\lambda  (\p^\mu  \omega^{\eta\nu} -
 \p^\nu \omega ^{\eta \mu })]+\mbox{}\nonumber\\
 && \mbox{}-\frac{1}{3}  \varepsilon ^{\mu \nu \alpha \beta }
 (\Box \omega ^\lambda_\lambda - \p_\lambda \p_\eta
 \omega^{\lambda\eta}); 
 \label{2.4}
 \end{eqnarray}
 \begin{eqnarray}
 \delta K^{\mu \nu \alpha \beta } &=&
 g^{\mu \alpha }(\p^\nu  \eta^\beta + \p^\beta  \eta^\nu )
 + g^{\nu \beta }(\p^\mu  \eta^\alpha  + \p^\alpha  \eta^\mu )
 - g^{\mu \beta } (\p^\nu  \eta^\alpha  + \p^\alpha   \eta^\nu  ) +
 \mbox{} \nonumber\\
 &&\mbox{} - g^{\nu \alpha }(\p^\mu  \eta ^\beta + \p^\beta \eta^\mu)
 - 2(g^{\mu \alpha } g^{\nu\beta} - g^{\mu \beta } g^{\nu \alpha })
 {\p_\sigma  \eta ^\sigma}, 
 \label{2.5}
 \end{eqnarray}
where the gauge tensor $\omega ^{\alpha \beta }$ is symmetric
$\omega^{\alpha \beta } = \omega ^{\beta \alpha }$.
 Not all components of $\omega^{\alpha\beta}$ act effecively
 because the transformation (\ref{2.4}) is invariant under
 \[
 \delta \omega^{\alpha\beta} = \p^\alpha \lambda^\beta +
 \p^\beta \lambda^\alpha
 \]
where $\lambda^\alpha$ is an arbitrary vector.  
We note that the transformation (\ref{2.4}) varies some
components of the Weyl part of $K^{\mu \nu \alpha \beta }$,
and the transformation (\ref{2.5}) varies some components of the
other parts of the field $K^{\mu \nu \alpha \beta }$.
The action integral describing the interaction with the external Weyl
current \\
 \be
 I_{int} = \int d^4 x {\cal L}_{int}
 \ee
is invariant under the gauge transformation (\ref{2.4}) if
 the source obeys the following condition
 \be
 \varepsilon ^{\alpha \lambda \mu \nu }\p_\lambda  \p_\sigma
 {j^{\sigma \beta }}_{\mu \nu } = 0 
 \label{2.7}
 \ee
where the dual properties of the Weyl tensor are taken into account.
Using the decomposition of the Weyl tensor (see Appendix II)
\[
j^{\mu\nu\alpha\beta} = (\lambda^{ij}, \sigma^{ij})
\]
we can rewrite the conservation law (\ref{2.7}) in the form 
 \be
 \begin{array}{l}
 \p_i \p_j \sigma^{ij} = 0, \nonumber\\
 \p^0 \sigma^i + \varepsilon^{ikp} \p_k \lambda_p  = 0,  \\ 
 \mbox{}[{(\p^0)}^2 + \Delta ] \sigma^{ij} + \p^i \sigma^j +
 \p^j \sigma^i +
 \p^0 [\varepsilon^{ikp} \p_k \lambda^j_p + \varepsilon^{jkp}
 \p_k \lambda^i_p] = 0, \nonumber
 \end{array}
 \ee
where
\[
  \sigma^i \equiv \p_j \sigma^{ji},\qquad
 \lambda^i \equiv \p_j \lambda^{ji}.
 \]
In the helicity components (see Appendix IV) we get
 \be
 \begin{array}{l}
 \sigma_L = 0, \nonumber \\
 \p^0 \sigma^i_T + \varepsilon^{ikp} \p_k \lambda_{Tp} = 0, \\   [0pt]
 [{(\p^0)}^2 + \Delta] {\sigma{ij}}(\pm 2) +  \p^0 [\varepsilon^{ikp}
 \p_k {\lambda^j_p}(\pm 2) + \varepsilon^{jkp} \p_k
 {\lambda^i_p}(\pm 2)] = 0.\nonumber
 \label{2.9}
\end{array}
\ee
The conservation law (\ref{2.7}) can be obtained as well from the field
equation following from the variational principle $\delta I = 0$.
We do not write down this field equation. For futher aim we
write the field equation in the covariant gauge
\begin{eqnarray}
&&{K^{\mu\nu}}_{\mu\nu}=0,\nonumber\\
&&\p_\mu {K^{\mu\nu\alpha}}_\nu = 0,\label{2.10}\\
&&\p_\mu K^{\mu\nu\alpha\beta} = - {1\over2}(\p^\alpha
{{K^{\mu\nu}}_\mu}^\beta - \p^\beta {{K^{\mu\nu}}_\mu}^\alpha)\nonumber  
\end{eqnarray}
It has the following form
\be
\Box C^{\mu\nu\alpha\beta} = 4 j^{\mu\nu\alpha\beta} 
\label{2.11}
\ee
We note that the gauge conditions (\ref{2.10}) lead to
\be
\varepsilon^{\alpha\lambda\mu\nu} \p_\lambda\p_\sigma
{C^{\sigma\beta}}_{\mu\nu} = 0.
\ee
So, the current conservation law (\ref{2.7}) follows from
 Eq.\ (\ref{2.11}). 

\section{Canonical analysis}

Using the decomposition of the Weyl tensor (see Appendix II)
 \[
 j^{\mu\nu\alpha\beta} = \left(\lambda^{ij},\sigma^{ij}\right)
 \]
and the Riemann tensor (see Appendix III)
 \[
   K^{\mu\nu\alpha\beta} = \left(T^{ij},R^{ij},S^{ij},A^{i},T,R\right)
 \]
we can rewrite the action (\ref{2.1}) in the component form. After some
integrations by parts we remove  the velocities $\p^0  A^i$, $\p^0 S^{ij}$
and $\p^0 R$ from the action. Performing the Legendre transformation
we  obtain
 \be
 I = \int d^4 x (P^{ij} \p^0 T_{ij} + \Pi^{ij} \p^0 R_{ij} + P \p^0 T -
 {\cal H}_c),
 \ee
where the canonical momenta are
 \begin{eqnarray}
   P^{ms} \equiv \frac{\p {\cal L}}{\p \p^0 T_{ms}} &=& - \left\{\p^0
   T^{ms} + {1\over2} \p^0 R^{ms} + {3\over2} (\p^s A^m + \p^m A^s)+
   \right.\nonumber \\
  &&\left. -  g^{ms} \p_k A^k - {1\over2} [\varepsilon^{mnp} \p_n S^s_p +
  \varepsilon^{snp} \p_n S^m_p ]\right\},  \nonumber\\
  \Pi^{ms} \equiv \frac{\p{\cal L}}{\p \p^0 R_{ms}} &=& - \left\{{1\over2}
  \p^0 T^{ms} + {1\over2} (\p^s A^m + \p^m A^s ) - {1\over3} g^{ms}
  \p_k A^k+ \mbox{}\right.\nonumber\\
  &&\left.- {1\over2} [\varepsilon^{mnp} \p_n S^s_p + \varepsilon^{snp}
  \p_n S^m_p]\right\},\\
  P \equiv \frac{\p {\cal L}}{\p \p^0 T} &=& {1\over3} \p^0 T - {4\over3}
  \p_k A^k,  \nonumber
 \end{eqnarray}
and the canonical Hamiltonian density is
 \begin{eqnarray}
  {\cal H}_c &=& 2 (\Pi^{ij})^2 - 2 \Pi_{ij} P^{ij} + {3\over2} P^2
   + {1\over6} T^i \p_i T + {1\over18} (\p^i T)^2 +
  {1\over2} (\p^k R^{ij})^2  + \mbox{}\nonumber\\
  &&\mbox{}+{1\over2} \p^k R^{ij} \p_k T_{ij}
  - (R^i)^2 - R^i T_i - {1\over6} R^i \p_i T - \lambda_{ij}
  (T^{ij} - R^{ij})+\mbox{}\nonumber\\
  &&\mbox{}+ 2 (P^i + \Pi^i - 2 \p^i P) A_i - ({1\over9} \Delta T -
  {1\over12} \p_i R^i - {1\over4} \p_i T^i) R+\mbox{}\nonumber\\
  &&\mbox{}+2[\varepsilon^{pnm} \p_n
  (P^s_m - \Pi^s_m) + \sigma^{ps}] S_{ps},
 \end{eqnarray}
where the following abbreviations are introduced \\
  $ T^i \equiv \p_j T^{ji}$, $R^i \equiv \p_j R^{ji}$,$P^i \equiv \p_j
  P^{ji}$, $\Pi^i \equiv \p_j \Pi^{ji}$.
The momenta conjugated to $A^i$, $S^{ij}$ and $R$ vanish because the
Lagrangian density is independent of the corresponding velocities
 \[
 p^{i}_{A} \equiv \frac{\p{\cal L}}{\p \p^0 A_i} = 0,\quad
 p^{ij}_{S} \equiv \frac{\p{\cal L}}{\p \p^0 S_{ij}} = 0,\quad
 p_R \equiv \frac{\p{\cal L}}{\p \p^0 R} = 0.
 \]
So, there are the following primary constraints
 \be
 \Phi^i_{(1)} = p^i_A,\qquad
 \Phi^{ij}_{(2)} = p^{ij}_S,\qquad
 \Phi_{(3)} = p_R.
 \ee
We introduce the total Hamiltonian [8]
 \be
 H_{tot} = \int d^3 x ({\cal H}_c + \lambda_i \Phi^i_{(1)} +
 \lambda_{ij} \Phi^{ij}_{(2)} + \lambda \Phi_{(3)}),
 \ee
where $\lambda_i$, $\lambda_{ij}$ and $\lambda$ are Lagrange
multipliers. The dynamics is expressed by
 \be
 \p^0 a = \left\{ a, H_{tot}
 \right\}_{\Phi_{(1)}=\Phi_{(2)}=\Phi_{(3)}=0},
 \ee
where $\left\{\ldots,\ldots\right\}$ is the Poisson bracket
and $a$ is a function of dynamical variables.
The theory is consistent if constraints hold for all times.
This leads to the secondary constraints:
 \begin{eqnarray}
 \Phi^i_{(4)} & = & P^i + \Pi^i - 2 \p^i P, \nonumber  \\
 \Phi^{ij}_{(5)} & = & \left[\varepsilon^{inp} \p_n \left(P^j_p
  - \Pi^j_p\right) +
 \varepsilon^{jnp} \p_n \left(P^i_p - \Pi^i_p\right)\right]
  + 2 \sigma^{ij},  \nonumber \\
  \Phi_{(6)} & = & \Delta T - {3\over4} \p_i R^i - {9\over4} \p_i T^i,
  \label{3.7} \\
  \Phi^{ij}_{(7)} & = & \varepsilon^{inm} \p_n
  \left[\Delta\left(T^j_m + R^j_m\right)
  + \p^j \left(T_m + R_m\right)\right]+\mbox{} \nonumber \\
  &&\mbox{}+\varepsilon^{jnm} \p_n \left[\Delta \left(T^i_m + R^i_m\right)
  + \p^i \left(T_m + R_m\right)\right]+\mbox{} \nonumber \\
  &&\mbox{}+ 4 \left[\p^0 {\sigma^{ij}} + \varepsilon^{inm}
  \p_n {\lambda^j_m}
  + \varepsilon^{jnm} \p_n {\lambda^i_m}\right],
  \nonumber
 \end{eqnarray} 
 where the conservation law (\ref{2.7}) is taken into account. We note
that
 \begin{eqnarray}
 \Phi^{ij}_{(5)} & = & \Phi^{ij}_{(5)}(\pm2)+\Phi^{ij}_{(5)}(\pm1),
 \nonumber\\
 \Phi^{ij}_{(7)} & = & \Phi^{ij}_{(7)}(\pm2),
 \label{3.8}
 \end{eqnarray}
because $\p_i\p_j\Phi^{ij}_{(5)}=0$ and $\p_j\Phi^{ij}_{(7)}=0$.
The dynamics of the constraints is
 \begin{eqnarray}
 \p^0 \Phi^i_{(1)} & = & - 2 \Phi^i_{(4)},  \nonumber \\
 \p^0 \Phi^{ij}_{(2)} & = & - \Phi^{ij}_{(5)}, \nonumber \\
 \p^0 \Phi_{(3)} & = & {1\over9} \Phi_{(6)}, \nonumber \\
 \p^0 \Phi^i_{(4)} & = & {2\over9} \p^i \Phi_{(6)},\nonumber \\
 \p^0 \Phi_{(6)} & = & {3\over2} \p_i \Phi^i_{(4)},\label{3.9}  \\
 \p^0 \Phi^{ij}_{(5)} & = & {1\over2} \Phi^{ij}_{(7)}, \nonumber \\
 \p^0 \Phi^{ij}_{(7)} & = & - 2 (\Delta \Phi^{ij}_{(5)} +
 \p^i \p_k \Phi^{kj}_{(5)} + \p^j \p_k \Phi^{ki}_{(5)}). \nonumber
  \end{eqnarray}
Because the constraints can be added to Hamiltonian [9] ,
we construct the new Hamiltonian density \\
 \be
 {\cal H}^{new} = {\cal H}_0 + V^{(4)}_i \Phi^i_{(4)} +
 V^{(5)}_{ij} \Phi^{ij}_{(5)} + V^{(6)} \Phi_{(6)} +
 V^{(7)}_{ij} \Phi^{ij}_{(7)},
 \ee
where $V^{(4)}_i$, $V^{(5)}_{ij}$, $V^{(6)}$ and $V^{(7)}_{ij}$
are the Lagrange multipliers, and ${\cal H}_0$ has the following form
 \begin{eqnarray}
 {\cal H}_0 & = & 2{(\Pi^{ij})}^2 - 2 \Pi_{ij} P^{ij} + {3\over2} P^2
 + {1\over6} T^i \p_i T+\mbox{} \nonumber \\
 &&\mbox{} + {1\over18}{(\p^i T)}^2 + {1\over2} {(\p^k R^{ij})}^2
 + {1\over2} \p^k R^{ij} \p_k T_{ij}+\mbox{} \\
 &&\mbox{} - {(R^i)}^2 - R^i T_i - {1\over6} R^i \p_i T - \lambda_{ij}
 (T^{ij} - R^{ij}). \nonumber
 \end{eqnarray}
We observe that the variables $A^i$, $S^{ij}$ and $R$ disappear
in the new description. Let us note that according to (\ref{3.8}) we have
(see Appendix IV)
 \begin{eqnarray*}
 V^{(5)}_{ij} & = & {V^{(5)}_{ij}}(\pm 2) + {V^{(5)}_{ij}}(\pm 1), \\
 V^{(7)}_{ij} & = & {V^{(7)}_{(ij)}}(\pm 2).
 \end{eqnarray*}
The Hamiltonian density ${\cal H}^{new}$ is derivable [10] from
the phase-space Lagrangian density
 \be
 {\cal L}^{new} = P_{ij} \p^0 T^{ij} + \Pi_{ij} \p^0 R^{ij} + P \p^0 T -
 {\cal H}^{new},
 \ee
where the Lagrangian multipliers $V^{(4)}_i$, $V^{(5)}_{ij}$,
$V^{(6)}$ and
$V^{(7)}_{(ij)}$ are treated as dynamical variables. So, passing to
the canonical formalism,  we find the primary constraints
 \begin{eqnarray}
 \pi^i_{(4)} = 0, \quad \pi^{ij}_{(5)} = 0,\quad
 \pi_{(6)} = 0, \quad \pi^{ij}_{(7)} = 0,
 \end{eqnarray}
where $\pi^i_{(4)}$, $\pi^{ij}_{(5)}$, $\pi_{(6)}$ and $\pi^{ij}_{(7)}$
are the canonical momenta conjugated to $V^{(4)}_i$, $V^{(5)}_{ij}$,
$V^{(6)}$ and $V^{(7)}_{ij}$ respectively. Thus the new total
Hamiltonian is
 \be
H^{new}_{tot} = \int d^3 x ({\cal H}^{new} + \lambda^{(4)}_i
 \pi^i_{(4)}) + \lambda^{(5)}_{ij} \pi^{ij}_{(5)} + \lambda^{(6)}
 \pi_{(6)} + \lambda^{(7)}_{ij} \pi^{ij}_{(7)}),
 \ee
where $\lambda^\prime$s are the Lagrange multipliers.
The dynamics is expressed by
 \be
 \p^0 a = \left\{ a ,H^{new}_{tot} \right\}_{\pi = 0}.
 \label{3.15}
 \ee
In particular we have
 \begin{eqnarray}
 \p^0 \pi^i_{(4)} = - \Phi^i_{(4)}, &&  \p^0 \pi^{ij}_{(5)} =
 -\Phi^{ij}_{(5)}, \nonumber \\
 \p^0 \pi_{(6)} = - \Phi_{(6)}, &&  \p^0 \pi^{ij}_{(7)} =
 - \Phi^{ij}_{(7)}.
 \end{eqnarray}
The time derivatives of $\Phi_{(4)}$, $\Phi_{(5)}$, $\Phi_{(6)}$
and $\Phi_{(7)}$ are given by Eqs.\ (\ref{3.9}).

\section{Gauge transformations}

Let us discuss the gauge transformations of the free notivarg
theory described by the action integral
 \be
 I_{free} = \int d^4 x {\cal L}^{new}_{free},
 \ee
where ${\cal L}^{new}_{free}$ is obtained from ${\cal L}^{new}$
putting $\lambda^{ij} = \sigma^{ij} = 0$.
In this limit the constraints are
 \begin{eqnarray}
 \Phi^i_{(4)} & = & P^i + \Pi^i - 2 \p^i P, \nonumber \\
 \Phi^{ij}_{(5)} & = & \varepsilon^{inp} \p_n (P^j_p - \Pi^j_p) +
 \varepsilon^{jnp} \p_n (P^i_p - \Pi^i_p), \nonumber \\
 \Phi_{(6)} & = & \Delta T - {3\over4} \p_i R^i - {9\over4}\p_i T^i, \\
 \Phi^{ij}_{(7)} & = & \varepsilon^{inm} \p_n [\Delta (T^j_m + R^j_m)
 + \p^j (T_m + R_m)]+\mbox{} \nonumber \\
 && + \varepsilon^{jnm} \p_n [\Delta (T^i_m + R^i_m) + \p^i (T_m + R_m)],
 \nonumber
 \end{eqnarray}
and they obey the relations
 \[
 \left\{ \Phi_{(a)} , \Phi_{(b)} \right\} = 0, \qquad
 a, b = 4, 5, 6, 7.
 \]
So, we have the theory with the first class constraints.\\
The generator of the gauge transformations is
 \begin{eqnarray}
 G = \int d^3 x \left(\alpha^{(4)}_{i}  \pi^{i}_{(4)} +
 \alpha^{(5)}_{ij}
 \pi^{ij}_{(5)} + \alpha^{(6)} \pi_{(6)} + \alpha^{(7)}_{ij}
 \pi^{ij}_{(7)}\right.\label{4.3}  \\
 \left.+\eta^{(4)}_{i} \Phi^{i}_{(4)} +
 \eta^{(5)}_{ij} \Phi^{ij}_{(5)}+
 \eta^{(6)} \Phi_{(6)} + \eta^{(7)}_{ij}
 \Phi^{ij}_{(7)}\right), \nonumber
 \end{eqnarray}
where $\alpha^\prime$s and $\eta^\prime$s are gauge
functions. They have the helicity
structure as the corresponding constraints.
The generator (\ref{4.3}) obeys the consistency condition
 \begin{eqnarray}
 \left.\frac{d}{dt} G \right|_{\pi = 0} = 0.
 \end{eqnarray}
Using Eq.\ (\ref{3.15}) we obtain
 \begin{eqnarray}
 \alpha^{(4)}_i & = & \p^0 \eta^{(4)}_i - {3\over2} \p_i \eta^{(6)},
\nonumber \\
 \alpha^{(6)} & = & \p^0 \eta^{(6)} - {2\over9} \p^i \eta^{(4)}_{i}, \\
 {\alpha^{(5)}_{ij}}(\pm 2) & = & \p^0 {\eta^{(5)}_{ij}}(\pm 2) -
 2 \Delta {\eta^{(7)}_{ij}}(\pm 2), \nonumber \\
 {\alpha^{(5)}_{ij}}(\pm 1) & = & \p^0 {\eta^{(5)}_{ij}}(\pm 1),
\nonumber \\
 \alpha^{(7)}_{ij} & = & \p^0 \eta^{(7)}_{ij} + {1\over2} \eta^{(5)}_{ij}
 (\pm 2). \nonumber
 \end{eqnarray}
The gauge transformations are
 \begin{eqnarray}
 \delta V^{(4)}_i & \equiv & \left\{ V^{(4)}_i, G \right\}
 = \alpha^{(4)}_i, \nonumber \\
 \delta V^{(5)}_{ij} & = & \alpha^{(5)}_{ij}, \nonumber \\
 \delta V^{(6)} & = & \alpha^{(6)}, \nonumber \\
 \delta V^{(7)}_{ij} & = & \alpha^{(7)}_{ij}, \nonumber \\
 \delta T & = & 2 \p^i \eta^{(4)}_{i}, \nonumber \\
 \delta P & = & - \Delta \eta^{(6)}, \nonumber \\
 \delta T^{ij} & = & - {1\over2} \left(\p^i \eta^{(4)j} +
 \p^j \eta^{(4)i}\right) +
 {1\over3} g^{ij} \p^k \eta^{(4)}_k+\mbox{}  \nonumber \\
 &&\mbox{}+\varepsilon^{inm} \p_n \eta^{(5)j}_m +
 \varepsilon^{jnm} \p_n  \eta^{(5)i}_m, \label{4.6} \\
 \delta R^{ij} & = & - {1\over2} \left(\p^i \eta^{(4)j} +
 \p^j \eta^{(4)i}\right) +
 {1\over3} g^{ij} \p^k \eta^{(4)}_k+\mbox{} \nonumber \\
 &&\mbox{}- \left[\varepsilon^{inm} \p_n \eta^{(5)j}_m +
 \varepsilon^{jnm} \p_n \eta^{(5)i}_m\right], \nonumber  \\
 \delta P^{ij} & = & {9\over4} \left(\p^i \p^j + {1\over3} g^{ij}
 \Delta\right)\eta^{(6)} - \Delta \left[\varepsilon^{inm} \p_n
 \eta^{(7)j}_m +
 \varepsilon^{jnm} \p_n \eta^{(7)i}_m\right], \nonumber \\
 \delta \Pi^{ij} & = & {3\over4} \left(\p^i \p^j +
 {1\over3} g^{ij} \Delta\right)
 \eta^{(6)} - \Delta \left[\varepsilon^{inm} \p_n \eta^{(7)j}_m +
 \varepsilon^{jnm} \p_n \eta^{(7)i}_m\right]. \nonumber
  \end{eqnarray}
In  Appendix V we impose the noncovariant conditions to remove
completely the gauge freedom.\\
Using the conservation law of the current (\ref{2.7}) we can verify
that the action
 \be
 I = \int d^4 x {\cal L}^{new}
 \label{4.7}
 \ee
is invariant under the gauge transformations (\ref{4.6}).

\section{Physical Lagrangian}

Solving the constraints (\ref{3.7}) we obtain \\
from $\Phi_{(4)} = 0$:
 \begin{eqnarray*}
  P_L + \Pi_L + 2 P = 0,  \\
 P^{i}_{T} + \Pi^{i}_{T} = 0,
 \end{eqnarray*}
from $\Phi_{(5)} = 0$:
\begin{eqnarray}
 P^{i}_{T} - \Pi^{i}_{T} & = & \frac{2}{\Delta} \varepsilon^{ikp}
 \p_k \sigma_{Tp}, \nonumber \\
 P^{ij}(\pm 2) - \Pi^{ij}(\pm 2) & = & \frac{1}{2 \Delta}
 \left[\varepsilon^{ikp} \p_k {\sigma^j_p}(\pm 2) +
 \varepsilon^{jkp} \p_k {\sigma^i_p}(\pm 2)\right],
 \end{eqnarray}
from $\Phi_{(6)} = 0$:
 \[
 T - {9\over4} T_L - {3\over4} R_L = 0,
 \]
from $\Phi_{(7)} = 0$:
 \[
 T^{ij}(\pm 2) + R^{ij}(\pm 2) = \frac{1}{ \Delta^2} \p^0
 [\varepsilon^{ikp}
 \p_k {\sigma^j_p}(\pm 2) + \varepsilon^{jkp} \p_k {\sigma^i_p}(\pm 2)]
 -\frac{4}{\Delta} {\lambda ^{ij}}(\pm 2).
 \]
Inserting these solutions to the action (\ref{4.7}) we get 
 \begin{eqnarray}
 I = \int d^4 x {\cal L}_{phys},
 \end{eqnarray}
where
 \begin{eqnarray}
 {\cal L}_{phys} = p \p^0 \varphi - {\cal H}_{free} - {\cal H}_{int}
 \end{eqnarray}
and
 \be
 (p , \varphi) = \left( \frac{\sqrt{3}}{2} (P_L - 3 \Pi_L) ,
 \frac{\sqrt{3}}{4} (T_L -R_L) \right),
 \label{5.4}
 \ee
or other 8 pairs that can be obtained from (\ref{5.4}) using the constraints:
 \[
 P_L + \Pi_L + 2 P = 0,\quad \mbox{\rm and} \quad
 T - {9\over4} T_L -{3\over4} R_L = 0.
 \]
The free Hamiltonian density is
 \begin{eqnarray}
 {\cal H}_{free} = {1\over2} p^2 - {1\over2} {(\p^i \varphi)}^2
 \end{eqnarray}
and the interacting one is
 \begin{eqnarray}
 {\cal H}_{int} & = & - 2 \sqrt{3} \lambda_L \varphi +
 4 {\lambda_{ij}}(\pm 2) \frac{1}{\Delta} {\lambda^{ij}}(\pm 2) +\mbox{}
 \label{5.6}\\
 &&\mbox{}+(\p^0 {\sigma^{ij}}(\pm 2))
 \frac{1}{\Delta^2} (\p^0 {\sigma_{ij}}(\pm 2))  -
 3 {\sigma^{ij}}(\pm 2) \frac{1}{\Delta} {\sigma_{ij}}(\pm 2) -
 8 \sigma^i_T \frac{1}{\Delta^2} \sigma_{Ti}. \nonumber
 \end{eqnarray}
In the momentum space we get
 \begin{eqnarray}
 {\cal H}_{int} & = & -2 \sqrt{3} {\lambda _L}(- k) \varphi (k)  +
 k^2 {(k^0)}^{-2} {\mid \vec{k} \mid}^{-2}
 {\sigma^{ij}}{({\pm 2} , {-k})} {\sigma_{ij}}{(\pm 2 , k)} + \mbox{}
 \nonumber \\
 &&\mbox{}+8 {\mid\vec{k}\mid}^{-4}
 {\sigma^i_T}(-k) {\sigma_{Ti}}(k). 
 \label{5.7}
 \end{eqnarray}

\section{Notivarg propagator}

Let us consider the exchange of the notivarg between two
external currents. The genaral structure of the amplitude
describing the process in the second order of the perturbation
theory is [7] 
 \begin{eqnarray}
 {\cal A} & = & - \left(\frac{a}{k^2} j^{\mu\nu\alpha\beta}(-k)
 j_{\mu\nu\alpha\beta}(k)+\frac{b}{k^4}k_\mu
 j^{\mu\nu\alpha\beta}(-k)k^\sigma j_{\sigma\nu\alpha\beta}+
 \mbox{}\right.\nonumber\\
 &&\left.\mbox{}+\frac{c}{k^6}k_\mu k_\alpha j^{\mu\nu\alpha\beta}(-k)
 k^\sigma k^\kappa j_{\sigma\nu\kappa\beta}(k)\right),
 \label{6.1}
 \end{eqnarray}
where $a$, $b$, $c$ are number factors. Due to the conservation law
(\ref{2.7}) we have
 \[
 k_\mu k_\alpha j^{\mu\nu\alpha\beta}
 k^\sigma k^\kappa j_{\sigma\nu\kappa\beta}=
 \frac{1}{2}k^2 k_\mu j^{\mu\nu\alpha\beta} k^\sigma
 j_{\sigma\nu\alpha\beta}.
 \]
Using the following identity for the Weyl tensor
 \[
 j^{\mu\nu\alpha\beta}j_{\sigma\nu\alpha\beta}=\frac{1}{4}
 \delta^{\mu}_{\sigma} j^{\kappa\nu\alpha\beta}
 j_{\kappa\nu\alpha\beta}
 \]
we observe only the first term in Eq.\ (\ref{6.1}) is independent.
Assuming the following form of the notivarg propagator
 \begin{eqnarray}
 {D_{\mu\nu\alpha\beta,\sigma\lambda\gamma\delta}}(k) & = &
 - {1\over8} \frac{1}{k^2} \left(g_{\mu\sigma} g_{\nu\lambda}
 g_{\alpha\gamma}
 g_{\beta\delta} + g_{\mu\lambda} g_{\nu\sigma} g_{\alpha\delta}
 g_{\beta\sigma} + \mbox{}\right. \nonumber \\
 &&\mbox{}+g_{\mu\gamma} g_{\nu\delta} g_{\alpha\sigma}
 g_{\beta\lambda} + g_{\mu\delta} g_{\nu\gamma} g_{\alpha\lambda}
 g_{\beta\sigma} - g_{\mu\lambda} g_{\nu\sigma} g_{\alpha\gamma}
 g_{\beta\delta} +\mbox{} \nonumber \\
 &&\left.\mbox{}-g_{\mu\sigma} g_{\nu\lambda} g_{\alpha\delta}
 g_{\beta\gamma} - g_{\mu\gamma} g_{\nu\delta} g_{\alpha\lambda}
 g_{\beta\sigma} - g_{\mu\delta} g_{\nu\gamma} g_{\alpha\sigma}
 g_{\beta\lambda}\right)
 \label{6.2}
 \end{eqnarray}
we obtain the amplitude
 \be
 {\cal A} = - j^{\mu\nu\alpha\beta} (-k) D_{\mu\nu\alpha\beta,
 \sigma\lambda\gamma\delta}(k) j^{\sigma\lambda\gamma\delta}(k)
 \ee
The number factor $a=1$ follows from Eqs (\ref{2.2}), (\ref{2.11}),
and (\ref{6.2}) 
 \[
 \frac{1}{4}j^{\mu\nu\alpha\beta}C_{\mu\nu\alpha\beta} \rightarrow 
 \frac{1}{4}j^{\mu\nu\alpha\beta}
 D_{\mu\nu\alpha\beta,\sigma\lambda\gamma\delta}
 \left(4 j^{\sigma\lambda\gamma\delta}\right).
 \]
Using the current conservation law (\ref{2.9}) we obtain 
 \begin{eqnarray}
 {\cal A} & = & 12 \frac{{\lambda _L}(-k){\lambda_L}(k)}{k^2} + 2
 k^2 {(k^0)}^{-2} {\mid\vec{k}\mid}^{-2} {\sigma^{ij}}(\pm 2 , -k)
 {\sigma_{ij}} ({\pm 2} , k) +\mbox{} \label{6.4} \\
 &&\mbox{}+16 {\mid\vec{k}\mid}^{-4} {\sigma^i_T}(-k) {\sigma_{Ti}}(k).
 \nonumber
  \end{eqnarray}
We note that the amplitude of the current - current interaction via 
one notivarg exchange can be calculated with the help of the Hamiltonian
(\ref{5.7}) using standard methods of the S - matrix formalism [11].
Following this way we get exactly the amplitude (\ref{6.4}). So, 
the form (\ref{6.2}) of the notivarg propagator is confirmed.

\section{Final remarks}

We finish with the following remarks:\\
(i) substituting $\lambda\rightarrow\frac{1}{2\sqrt{2}}\sigma$,
$\sigma\rightarrow-\frac{1}{2\sqrt{2}}\lambda$
(see Appendix II) in Eq.\ (\ref{5.6})
we get the physical Hamiltonian in the notivarg theory based on the
Lagrangian (\ref{1.2}) (see Refs \ [6,7]);\\
(ii) in the Deser-Siegel-Townsend description the scalar field is not
necessarily a component of the Weyl tensor. So, we can expect the
notivarg interacting with other parts of the external Riemann current,
not only with the Weyl one.

\section{Acknowledgment}

We thank  Prof.\ J.\ Rembieli\'nski and Dr.\ J.\ K{\l}osi\'nski for
interesting discussion.

\section*{Appendix I}

The decomposition of the Riemann tensor $K^{\mu\nu\alpha\beta}$
in the irreducible Lorentz parts is:\\
 \[
 K^{\mu\nu\alpha\beta} = C^{\mu\nu\alpha\beta} +
 E^{\mu\nu\alpha\beta} + G^{\mu\nu\alpha\beta},
 \]
where  $C^{\mu\nu\alpha\beta}$ is the Weyl tensor and
 \begin{eqnarray*}
 E^{\mu\nu\alpha\beta} & = & {1\over2} (g^{\mu\alpha} K^{\nu\beta} +
 g^{\nu\beta} K^{\mu\alpha} - g^{\mu\beta} K^{\nu\alpha} -
 g^{\nu\alpha} K^{\mu\beta})+\mbox{}\\
 &&\mbox{}- {1\over4} (g^{\mu\alpha} g^{\nu\beta}
 - g^{\mu\beta} g^{\nu\alpha}) K,  \\ 
 G^{\mu\nu\alpha\beta} & = & \frac{1}{12} (g^{\mu\alpha} g^{\nu\beta}
 - g^{\mu\beta} g^{\nu\alpha}) K,  \\
 K^{\mu\alpha} & \equiv & {K^{\mu\nu\alpha}}_{\nu},\quad
 K \;\equiv\; K^\mu_\mu.
 \end{eqnarray*}
The dual property of the Weyl tensor reads
 \[
 {\varepsilon^{\mu \nu}}_{\sigma \lambda} C^{\sigma \lambda \alpha \beta}
 = {\varepsilon^{\alpha \beta}}_{\sigma \lambda}
 C^{\mu \nu \sigma \lambda}.
 \]

\section*{Appendix II}

Let us consider the Weyl tensor $j^{\mu\nu\alpha\beta}$. We introduce 
the new variables 
\[
\lambda^{ij} = \lambda^{ji},\quad \lambda^i_i = 0,\quad
 \sigma^{ij} = \sigma^{ji},\quad \sigma^i_i = 0
\]
defined by 
 \begin{eqnarray*}
 j^{0i0j} & = & \lambda^{ij},\\
 j^{0ijk} & = & \varepsilon^{jkp} \sigma^i_p, \\
 j^{ijkl} & = & - \left(g^{ik} \lambda^{jl} + g^{jl} \lambda^{ik} -
 g^{il} \lambda^{jk} - g^{jk} \lambda^{il}\right).
 \end{eqnarray*}
So, we get the following decomposition of the Weyl tensor 
\[
j^{\mu\nu\alpha\beta} = \left(\lambda^{ij}, \sigma^{ij}\right).
\]
The dual transformation 
\[
j^{\mu\nu\alpha\beta} \rightarrow \frac{1}{2} {\varepsilon^{\mu\nu}}_
{\sigma\lambda} j^{\sigma\lambda\alpha\beta}
\]
in the component form is 
\[
\lambda^{ij} \rightarrow - \sigma^{ij},\qquad \sigma^{ij} \rightarrow 
\lambda^{ij}.
\]

\section*{Appendix III}

Let us consider the Riemann tensor $K^{\mu\nu\alpha\beta}$.
We introduce the new variables
 \begin{eqnarray*}
 T^{ij} = T^{ji},\quad  T^i_i = 0,\quad
 R^{ij} = R^{ji},\quad  R^i_i = 0,  \\
 S^{ij} = S^{ji},\quad  S^i_i = 0,\quad  A^i,\quad  T,\quad  R
 \end{eqnarray*}
defined by
 \begin{eqnarray*}
 K^{0i0j} & = & T^{ij} + {1\over3} g^{ij} T,  \\
 K^{0ijk} & = & \varepsilon^{jkp} S^i_p + g^{ij} A^k - g^{ik} A^j,  \\
 K^{ijmn} & = &  g^{im} R^{jn} + g^{jn} R^{im} - g^{in} R^{jm} -
 g^{jm} R^{in}+\mbox{} \\
 &&\mbox{}+ {1\over6} (g^{im} g^{jn} - g^{in} g^{jm}) R.
 \end{eqnarray*}
So, we get the following decomposition of the Riemann tensor
 \[
 K^{\mu\nu\alpha\beta} = (T^{ij}, R^{ij}, S^{ij}, A^i, T, R).
 \]

\section*{Appendix IV}

The well known decomposition of a vector into transversal and
longitudinal parts is
 \[
 V^i = V^i_T + V^i_L
 \]
where
 \begin{eqnarray*}
 V^i_T & = & V^i + \frac{1}{\Delta} \p^i \p_j V^j,  \\
 V^i_L & =  & - \frac{1}{\Delta} \p^i \p_j V^j,  \\
 \Delta & = & - \p_i \p^i.
 \end{eqnarray*}
The analogous decomposition of a symmetric traceless tensor
$a^{ij}$ is
 \[
 a^{ij} = {a^{ij}}(\pm 2) + {a^{ij}}(\pm 1) + {a^{ij}}(0)
 \]
where
 \begin{eqnarray*}
 a^{ij}(\pm 1) & = & - \frac{1}{\Delta} (\p^i a^j_T + \p^j a^i_T),  \\
 a^{ij}(0) & = & {3\over2} (\frac{1}{\Delta} \p^i \p^j +
 {1\over3} g^{ij}) a_L, \\
 a^i_T & = & a^i + \frac{1}{\Delta} \p^i \p_j a^j, \\
 a_L & = & \frac{1}{\Delta} \p_i a^i, \\
 a^i & = & \p_j a^{ji}.
 \end{eqnarray*}

\section*{Appendix V}

The gauge transformations (\ref{4.6}) in the component form are
 \begin{eqnarray*}
 \delta {T^{ij}}(\pm 2) & = & - \delta {R^{ij}}(\pm 2) = \varepsilon^{ikp}
 \p_k {\eta^{(5)j}_p}(\pm 2) + \varepsilon^{jkp} \p_k
 {\eta^{(5)i}_p}(\pm 2),  \\
 \delta T^i_T & = & {1\over2} \Delta \eta^{(4)i}_T  + \varepsilon^{ikp}
 \p_k \eta^{(5)}_{Tp}, \\
 \delta R^i_T & = & {1\over2} \Delta \eta^{(4)i}_T  - \varepsilon^{ikp}
 \p_k \eta^{(5)}_{Tp}, \\
 \delta R_L & = & {2\over3} \Delta \eta^{(4)}_L,  \\
 \delta T_L & = & {2\over3} \Delta \eta^{(4)}_L,  \\
 \delta T & = & 2 \Delta \eta^{(4)}_L,  \\
 \delta {P^{ij}}(\pm 2) & = & \delta {\Pi^{ij}}(\pm 2) = - \Delta
 \left(\varepsilon^{ikp}
 \p_k {\eta^{(7)j}_p}(\pm 2) + \varepsilon^{jkp} \p_k
 {\eta^{(7)i}_p}(\pm 2)\right),  \\
 \delta P^i_T & = & \delta \Pi^i_T = 0,  \\
 \delta P_L & = & {3\over2} \Delta \eta^{(6)},  \\
 \delta \Pi_L & = & {1\over2} \Delta \eta^{(6)},  \\
 \delta P & = & - \Delta \eta^{(6)}.
 \end{eqnarray*}
To remove completely the gauge freedom we impose the following
noncovariant conditions  $\left\{\chi_i \right\}$: 
 \begin{eqnarray*}
 && {T^{ij}}(\pm 2) - {R^{ij}}(\pm 2) = 0,\qquad  T^i_T = R^i_T = 0,   \\
 && {P^{ij}}(\pm 2) + {\Pi^{ij}}(\pm 2) = 0,\qquad
  aT - \frac{3}{8}(3T_L + R_L) = 0,\\
 &&  P + a (P_L + \Pi_L) = 0,\qquad   a \neq \frac{1}{2}, \\
 && \left\{\chi_i,\chi_j \right\} = 0.
 \end{eqnarray*}
The constraints  $\Phi^\prime$s and the gauge condition
$\chi^\prime$s form the set of the second class constraints.

%%%%%%%%%%%%%%%%%%%%%%%%%%%%%%%%%%%%%%%%%%%%%%%%%%%%%%%%

\end{document}